\documentclass[conference]{IEEEtran}
\usepackage[scr=rsfs]{mathalpha}

\newcommand{\tabincell}[2]{\begin{tabular}{@{}#1@{}}#2\end{tabular}}

\usepackage{graphicx, amsfonts,amsmath , boldline, epstopdf,amssymb }
\usepackage{tabu, multirow,multicol ,booktabs ,setspace ,algcompatible ,mathrsfs,dsfont, array, boldline, makecell, booktabs }
\usepackage{amssymb}
\usepackage[linesnumbered,lined,boxed,commentsnumbered,ruled,longend]{algorithm2e}
\usepackage{adjustbox, stackengine, color, colortbl, epsfig,cite, cases, enumitem, mathtools, tikz, verbatim}
\newcolumntype{?}{!{\vrule width 1.2pt}} 
\usepackage[caption = false]{subfig}
\usepackage[utf8]{inputenc}
\makeatletter

\begin{document}

\title{\vspace{-.9cm} Security-Constrained Optimal Operation of Energy-Water Nexus based on a Fast Contingency Filtering Method \\
\thanks{Identify applicable funding agency here. If none, delete this.}
}

\author{\IEEEauthorblockN{ Mostafa Goodarzi}
\IEEEauthorblockA{\textit{Electrical and Computer Engineering} \\
\textit{University of Central Florida}\\
Orlando, USA \\
mostafa.goodarzi@knights.ucf.edu}
\and
\IEEEauthorblockN{ Qifeng Li}
\IEEEauthorblockA{\textit{Electrical and Computer Engineering} \\
\textit{University of Central Florida}\\
Orlando, USA \\
Qifeng.Li@ucf.edu}}

\maketitle

\begin{abstract}
Water and power systems are increasingly interdependent due to the growing number of electricity-driven water facilities. The security of one system can be affected by a contingency in the other system. This paper investigates a security-constrained operation problem of the energy-water nexus (EWN), which is a computationally challenging optimization problem due to the nonlinearity, nonconvexity, and size. We propose a two-step iterative contingency filtering method based on the feasibility and rating of the contingencies to decrease the size of the problem. The optimal power and water flow are obtained in a normal situation by considering the set of contingencies that can not be controlled with corrective actions. The feasibility check of the contingencies is performed in the second step, followed by a rating of the uncontrollable contingencies. Finally, the critical contingencies are obtained and added to the first step for the next iteration. We also employ convex technologies to reduce the computation burden. The proposed method is validated via two case studies. Results indicate that this approach can efficiently attain optimal values.
\end{abstract}

\begin{IEEEkeywords}
 contingency filtering, energy water nexus, optimal power and water flow, security constrained.
\end{IEEEkeywords}

\allowdisplaybreaks
\section{Introduction}
Water and power systems are closely linked \cite{li2018modeling}, and their interdependence is increasing because of the rising utilization of electricity-driven water facilities (EDWFs). On the one hand, water scarcity can be tackled by enhancing water efficiency through recycling, which heavily relies on electric power. On the other hand, EDWFs are able to adjust to power imbalances since their demand is flexible. Hence, the water distribution system (WDS) and the power distribution network (PDN) are better served by working together as an energy-water nexus (EWN).

Insufficient power supply to the EDWFs could lead to a failure in the PDN. Furthermore, a contingency in the WDS could alter the operation of the EDWFs and their power demand. These changes in the operation of the EDWFs have created an imbalance between power supply and demand, resulting in decreased PDN security. This paper proposes an N-1 security-constrained optimal operation of an EWN (SCOEWN) with corrective actions to address these contingency problems and increase system security.

The SCOEWN is a combination of the optimal power and water flow (OPWF) and the security-constrained optimal power flow (SCOPF), which is used for a period of time. The SCOPF and OPWF are obtained from large-scale, nonconvex, and nonlinear optimization problems that are computationally intractable and difficult to solve \cite{xavier2019transmission,park2018sparse}. Computing the SCOEWN is computationally challenging due to a large number of decision variables, nonlinearity, and nonconvexity. It would require prohibitive amounts of memory and CPU time to reach the SCOEWN directly.

In real-life applications, most contingencies have no impact on the optimal solution, and only a limited subset of contingencies is critical to the security optimal operation \cite{ardakani2013identification,weinhold2020fast}. In several studies, this principle has been applied to reduce the size of the security-constrained optimization problem by detecting uncontrollable contingencies (UCs) using contingency filtering methods \cite{capitanescu2008new,jiang2013novel,fliscounakis2013contingency}. All of these methods only consider the power network at the transmission level and a one-time slot. In this paper, we consider power distribution levels with radial structure, integrated systems, and the parallel computation framework to speed up computations in our proposed method.

To ensure the secure operation of EWN, we propose an iterative two-stage approach based on contingency rating along with simultaneous testing of feasibility (CRSFT). This method takes into account only the crucial contingencies when solving the optimization problem. To begin with, we find the optimal power and water flow for the EWN subject to the regular operation constraints and the set of critical contingencies. In the next step, the feasibility of the contingencies is checked by considering the corrective actions. The parallel process is applied for analyzing different optimization problems to speed up the computation \cite{gao2021benders}. The UCs, which can not be controlled by corrective action, are determined and rated by their violation of the feasible solution. Based on the contingency rating, the critical contingency is identified and added to the set of critical contingencies for the next iteration. The method will be terminated when all of the contingencies lead to a feasible solution by taking into account the corrective actions. In addition, we apply convex technologies to both stages of the proposed method to reduce the computational burden.

The rest of this paper is organized as follows. The details of the problem formulation are presented in section \ref{sec:PF}. Section \ref{sec:SolMet} introduces the solution method. Section \ref{sec:casestudy} presents two case studies to validate the CRSFT method. Lastly, conclusions are drawn in Section \ref{sec:conclusions}.

\section{Problem Formulation}  \label{sec:PF}
In this section, we formulate the security-constrained operation problem of an EWN system with corrective actions. An EWN includes a PDN, a WDS, and EDWFs that are modeled in the following subsections.

\subsection{Power Distribution Network} \label{sec:PDNnorml}
In this paper, we use the well-known $\mathrm{Distflow}$ model to describe power flow in the distribution system \cite{baran1989optimal}. Besides, distribution network reconfiguration (DNRC) is considered as a corrective action to avoid isolating all customers downstream of the on-contingency area in the radial structure. According to this model, the formulations are expressed by (\ref{eq_SCOPF_1}) to (\ref{eq_SCOPFRC_2}):
\begin{subequations} \label{eq_SCOPF}
	\begin{align}
&P_{i,t}^\mathrm{g}+{\Delta P}_{i,t}^{\mathrm{g},k}-P_{i,t}^\mathrm{l}= r_{ij}\mathcal{I}_{ij,t}^{k}-P_{ji,t}^{k}+\sum_{k} P_{ki,t}^{k}, \label{eq_SCOPF_1}\\
&Q_{i,t}^\mathrm{g}+{\Delta Q}_{i,t}^{\mathrm{g},k}-Q_{i,t}^\mathrm{l}=x_{ij}\mathcal{I}_{ij,t}^{k}-Q_{ji,t}^{k}+\sum_{k} Q_{ki,t}^{k}, \label{eq_SCOPF_2}\\
&\mathcal{V}_{i,t}^{k}-\mathcal{V}_{j,t}^{k}=2(r_{ij}P_{ij,t}^\mathrm{k}+x_{ij}Q_{ij,t}^{k})-z_{ij}^2\mathcal{I}_{ij,t}^{k},\label{eq_SCOPF_3} \\
&(P_{ij,t}^{k})^2+(Q_{ij,t}^{k})^2=\mathcal{V}_{i,t}^{k}\mathcal{I}_{ij,t}^{k},\label{eq_SCOPF_4}\\
&(P_{ij,t}^{k})^2+(Q_{ij,t}^{k})^2 \leq \overline{S_{ij}}^2,\label{eq_SCOPF_5}\\
&\underline{P_{i}^\mathrm{g}},\underline{Q_{i}^\mathrm{g}}\leq P_{i,t}^\mathrm{g},Q_{i,t}^\mathrm{g} \leq \overline{P_{i}^\mathrm{g}},\overline{Q_{i}^\mathrm{g}},\label{eq_SCOPF_6}\\
&\underline{\mathcal{V}_i},\underline{\mathcal{I}_{ij}}, \leq \mathcal{V}_{i,t}^{k},\mathcal{I}_{ij,t}^{k}  \leq \overline{\mathcal{V}_i}, \overline{\mathcal{I}_{ij}},\label{eq_SCOPF_7}\\
&\underline{\mathbb{P}_{i}^\mathrm{g}}, \underline{\mathbb{Q}_{i}^\mathrm{g}} \leq {\Delta P}_{i,t}^{\mathrm{g},k}, {\Delta Q}_{i,t}^{\mathrm{g},k} \leq \overline{\mathbb{P}_{i}^\mathrm{g}},\overline{\mathbb{Q}_{i}^\mathrm{g}}\label{eq_SCOPF_8}\\
&\begin{cases}
\!\mathcal{V}_{i,t}^\mathrm{k}\!-\!\mathcal{V}_{j,t}^{k}\!\!+\!\!z_{ij}\mathcal{I}_{ij,t}^\mathrm{k}\!=\!2(r_{ij}P_{ij,t}^\mathrm{k}\!+\!x_{ij}Q_{ij,t}^{k}\!),&\!\text{if}\! \ \!\alpha^k_{ij}\!=\!1\\
\!\mathcal{I}_{ij,t}^{k},P_{ij,t}^{k},Q_{ij,t}^{k}\! = \!\!0, \!&\text{if}  \alpha^k_{ij}\!=\!0 \label{eq_SCOPFRC_1}
\end{cases} \\
&\sum_{i,j} \alpha^\mathrm{k}_{ij}= N_b-1,\label{eq_SCOPFRC_2}
\end{align}
\end{subequations}
where $k$, $i$ and $j$ and $t$ show the number of contingency, the number of bus, and time, respectively; $k = 0$ shows the base case operation without any contingency. Equations (\ref{eq_SCOPF_1}) and (\ref{eq_SCOPF_2}) show the nodal balance of power, and (\ref{eq_SCOPF_3}) to (\ref{eq_SCOPF_5}) are related to Ohm's law. $P_{i,t}^\mathrm{g}$, $Q_{i,t}^\mathrm{g}$, $P_{i,t}^\mathrm{l}$, $Q_{i,t}^\mathrm{l}$, and $\mathcal{V}_{i,t}^\mathrm{k}$ are active and reactive power generation and power load and square voltage of the bus, respectively. $P^k_{ij,t}$ and $Q^k_{ij,t}$, $\mathcal{I}_{ij,t}^{k}$, $r_{ij}$, $x_{ij}$, and $z_{ij}$ are active and reactive power flow, the square of the current magnitude, resistance, reactance, and the sum of the square of reactance and resistance of the line, respectively. We consider the power generation changes as a corrective action in our model which are shown with ${\Delta P}_{i,t}^{\mathrm{g},k}$ and ${\Delta Q}_{i,t}^{\mathrm{g},k}$ that are limited to the acceptable value of ramping down and up ($\underline{\mathbb{P}_{i}^\mathrm{g}}$, $\underline{\mathbb{Q}_{i}^\mathrm{g}}$, $\overline{\mathbb{P}_{i}^\mathrm{g}}$, and $\overline{\mathbb{Q}_{i}^\mathrm{g}}$). Equations (\ref{eq_SCOPF_6}) to (\ref{eq_SCOPF_8}) describes the upper and lower bounds of the variables. The DNRC which is another corrective action is represented by (\ref{eq_SCOPFRC_1}) and (\ref{eq_SCOPFRC_2}). $\alpha^k_{ij}$ shows the status of a contingency for the line and $N_b$ is the total number of buses.

\subsection{Water Distribution System}
In this paper a model of WDS which consists of mass flow conservation law, pipe network, water tank, pressure-reducing valve (PRV), and water pump model is represented as follows:
\begin{subequations} \label{eq_SW}
	\begin{align}
&\sum\limits_{l}f^{k}_{nl,t} = F^\mathrm{R}_{n,t}+\hat{ F}^{\mathrm{R},k}_{n,t}-d_{n,t}+F^{\mathrm{T},k}_{n,t}, \label{eqSW_1}\\
& y^{k}_{n,t}-y^{k}_{l,t}=R_{nl}^{\mathrm{w}}sgn(f^{k}_{nl,t})(f^{k}_{nl,t})^2,\label{eq_pipenet_1} \\
&\begin{cases}
      y^\mathrm{k}_{n,t}-y^\mathrm{k}_{m,t}+y^{\mathrm{G},k}_{nl,t}=R_{nl}^{\mathrm{w}}(f^k_{nl,t})^2 &\text{if} \ \beta_{nl,t} = 1, \\
      f^k_{nl,t}=0 &\text{if} \ \beta_{nl,t} = 0,
    \end{cases}, \label{eq_pipenet_2}\\
&y_{n,t+1}^\mathrm{T,k}=y_{n,t}^\mathrm{T,k}+{\frac{F_{n,t}^{\mathrm{T},k}} {A_{n}^\mathrm{T}}}, \label{eq_tank_1}\\
&V_{n,t+1}^{\mathrm{T},k}=V_{n,t}^{\mathrm{T},k}+F_{n,t}^{\mathrm{T},k}, \label{eq_tank_2}\\
&V_{n,0}^{\mathrm{T},k}=V_{n,24}^{\mathrm{T},k},\label{eq_tank_3}\\
&\underline{F^\mathrm{R}_{n}}, \underline{F^\mathrm{T}_{n}},\underline{f_{nl}} \leq F^\mathrm{R}_{n,t}, F^{\mathrm{T},k}_{n,t},f^\mathrm{k}_{nl,t} \leq \overline{F^\mathrm{R}_{n}}, \overline{F^\mathrm{T}_{n}},\overline{f_{nl}},\label{eqSW_2}\\
&\underline{\mathbb{F}^\mathrm{R}_n} \leq \hat{ F}^{\mathrm{R},k}_{n,t} \leq \overline{\mathbb{F}^\mathrm{R}_n},\label{eqSW_3}\\
&\underline{y_n}, \underline{V_{n}^\mathrm{T}} \leq y^k_{n,t}, V_{n,t}^{\mathrm{T},k}\leq \overline{y_n},\overline{V_n^\mathrm{T}}, \label{eq_pipenet_3}\\
& 0 \leq R_{nl}^{\mathrm{w}} \leq \Phi_{nl}, \ \ nl \in \xi_{rp}, \label{eq_PRV_1}\\
&R_{nl}^{\mathrm{w}}= \frac{8f_\mathrm{s}{L_{nl}}}{\pi ^{2}{g}D_{nl}^5},  \ \ nl \in \xi \setminus \xi_{rp}, \label{eq_R} 
	\end{align}
\end{subequations}
where $F^\mathrm{R}_{n,t}$, $d_{n,t}$,  $y^{k}_{n,t}$, $F^{\mathrm{T},k}_{n,t}$, $V^{\mathrm{T},k}_{n,t}$, and $A^\mathrm{T}_k$ show water flow injected from the water source, water demand, water head, net water flow of tank, water tank volume, and water tank area for node $n$, respectively. $\hat{ F}^{\mathrm{R},k}_{n,t}$ represents the change in water production that is considered as a corrective action in our model. $f^{k}_{nl,t}$, $y^{\mathrm{G},k}_{n,t}$ and $\beta_{nl,t}$ show water flow, head gains imposed by the pump and the pump status for pipe $nl$. 
In this model, (\ref{eqSW_1}) guarantees the total water injection to each node is equal to the whole withdraw water. Equations (\ref{eq_pipenet_1}) and (\ref{eq_pipenet_2}) show the head lost along a regular pipe and a pipe with a pump, respectively. Equation (\ref{eq_tank_1}) describes the head pressure change at the water tank node; (\ref{eq_tank_2}) shows the stored water in the water tank at each time slot; (\ref{eq_tank_3}) represents that the total water input to the tank in a day should be equal to the total water output from the tank. 
Equations (\ref{eqSW_2}) to (\ref{eq_pipenet_3}) describe the upper and lower bounds of variables. There are several types of controllable valves in WDSs that can help operators to control the WDS operation \cite{fooladivanda2017energy}. In this paper, we assume that the system operators use PRV to reduce the water pressure at some specific pipes in order to control the water head pressure. Equation (\ref{eq_PRV_1}) shows the acceptable value for the head loss coefficient of the pipes with PRVs, where $\Phi_{nl}$ shows the maximum value of head loss in the set of pipes with PRVs ($\xi_{rp}$). Equation (\ref{eq_R}) shows the Darcy-Weisbach formula that is the most theoretically accurate formula to calculate the head lost along the pipe without any PRVs \cite{goodarzi2021fast}, where $L_{nl}$ and $D_{nl}$ are the length and diameter of the pipes, respectively, $f_s$ is the coefficient of surface resistance, and $g$ represents the gravitational acceleration.

\subsection{Electricity Driven Water Facilities}
The WDS and the PDN are linked by EDWFs, which require power and water. In this paper, a fixed-speed pump is considered as an EDWF and formulated by a quadratic function of the water flow as \cite{ulanicki2008modeling}:
\begin{align} \label{eq_pump}
\eta P^{\textrm {pump},k}_{i,t} &= 2.725\big[a_1{f^c_{nl,t}}^2+a_0f^c_{nl,t}\big].
\end{align}

\section{Solution Method}   \label{sec:SolMet}
This section explains the proposed method of solving SCOEWN. First, we discuss the convex relaxation for the EWN model to reduce the computational burden. Then, the CRSFT method is explained to decrease the problem size.

\subsection{Convex Relaxation} \label{sec:SolMetconv}
In this section, we convexify the formulation of the EWN to reduce the computational burden.
\subsubsection{Convexification of Constraint (\ref{eq_SCOPF_4})}
We relax (\ref{eq_SCOPF_4}) by using the convex hull relaxation model that is represented in \cite{li2018micro} as follows:
\begin{align}
&(P^k_{ij,t})^2+(Q^k_{ij,t})^2 \leq \mathcal{V}^k_{i,t}{\mathcal{I}^k}_{ij,t}, \label{ConvOPF1} \\
& \underline {\mathcal{V}_{i}}\overline{\mathcal{V}_{i}}{\mathcal{I}^k}_{ij,t}+ \overline{S_{ij}}^2\mathcal{V}^k_{i,t} \leq \overline{S_{ij}}^2(\underline {\mathcal{V}_{i}}+\overline{\mathcal{V}_{i}}). \label{ConvOPF2}
	\end{align}
\subsubsection{Convexification of Constraints (\ref{eq_SCOPFRC_1}) and (\ref{eq_pipenet_2})}
We apply the big-M technique to eliminate the logic proposition of (\ref{eq_SCOPFRC_1}) and (\ref{eq_pipenet_2}). Constraint (\ref{eq_SCOPFRC_1}) is replaced by (\ref{eq_ConvNRC_1}) and (\ref{eq_ConvNRC_2}). Besides, constraints (\ref{Convwatp1}) and (\ref{Convwatp2}) show the convex model of (\ref{eq_pipenet_2}). 
\begin{align}
& \resizebox{.9\hsize}{!}{$M\big(\alpha^k_{ij}\!-\!1\big)\! \le \!\mathcal{V}_{i,t}^k\!-\!\mathcal{V}_{j,t}^{k}\!+\!z_{ij}\mathcal{I}_{ij,t}^k  \!-\!2\big(r_{ij}P_{ij,t}^k\!+\!x_{ij}Q_{ij,t}^k\big)\! \le \!M\big(1\!-\!\alpha^c_{ij}\big)$}, \label{eq_ConvNRC_1}\\
&0 \leq {\mathcal{I}_{ij,t}}^{k} \leq \alpha^{k}_{ij}\overline{\mathcal{I}_{ij}}. \label{eq_ConvNRC_2}\\	
&\resizebox{.88\hsize}{!}{$M\big(\!\beta_{nl,t}\!-\!1\big) \!\le\! y^k_{n,t}\!-\!y^k_{l,t}\!+\!y^{\mathrm{G},k}_{nl,t}\!-\!R^\mathrm{w}_{nl}\overline{f}_{nl}f^k_{nl,t} \!\le\! M\big(\!1-\!\beta_{nl,t}\big)$},\label{Convwatp1}\\
&0 \le f^k_{nl} \le \beta^\mathrm{k}_{nl,t} \overline{f}_{nl}.\label{Convwatp2}
\end{align}
\subsubsection{Convexification of Constraint (\ref{eqSW_2})}
Equation (\ref{eqSW_2}) is a non-convex constraint that can be relaxed into convex hull relaxation mode as follow \cite{li2018micro}:
\begin{align}
&y^k_{n,t}-y^k_{l,t} 
\begin{cases}
      \!\leq 0.82R^\mathrm{w}_{nl}\overline{f}_{nl}f^k_{mn,t}+0.18R^\mathrm{w}_{nl}\overline{f}_{nl}^2 \\
      \!\geq 0.82R^\mathrm{w}_{nl}\underline{f}_{nl}f^k_{nl,t}-0.18R^\mathrm{w}_{nl}\underline{f}_{nl}^2  \\
      \!\geq 2R^\mathrm{w}_{nl}\overline{f}_{nl}f^k_{nl,t}-R^\mathrm{w}_{nl}\overline{f}_{nl}^2  \\
      \!\leq 2R^\mathrm{w}_{nl}\underline{f}_{nl}f^k_{nl,t}+R^\mathrm{w}_{nl}\underline{f}_{nl}^2
    \end{cases}\label{Convpipe}	
\end{align}

\subsubsection{Convexification of Constraint (\ref{eq_pump})}
A quadratic equation like (\ref{eq_pump}) is non-convex and it can be relaxed as the intersection of a concave inequality and a convex inequality:
	\begin{align}
&\eta P^{\textrm{pump},k}_{i,t} \ge 2.725\big(a_1{(f^k_{nl,t})}^2+a_0{f}^{k}_{nl,t}\big), \notag \\
&\eta P^{\textrm{pump},k}_{i,t}\le  2.725\big(a_1\overline {f}_{nl}+a_0\big) f^k_{nl,t}. \notag  
	\end{align}

\subsection{Contingency Rating with Simultaneous Feasibility Testing} \label{sec:SolMetiter}
\begin{figure}[!htb]
    \vspace{-.2cm}
  \centering
  \subfloat[]{\includegraphics[width=0.15\textwidth]{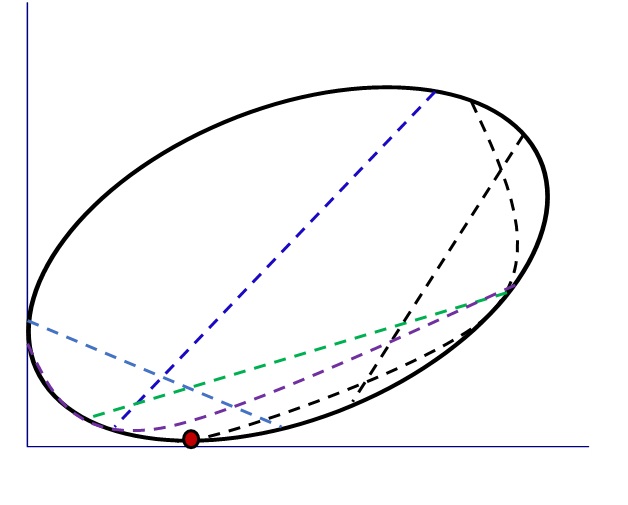} \label{Conv1}}
  \ \  \subfloat[]{\includegraphics[width=0.15\textwidth]{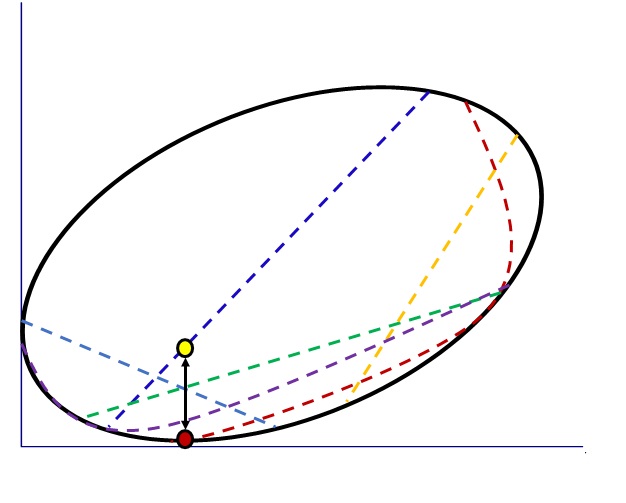}\label{Conv2}}\\
      \vspace{-.4cm}
    \subfloat[]{\includegraphics[width=0.15\textwidth]{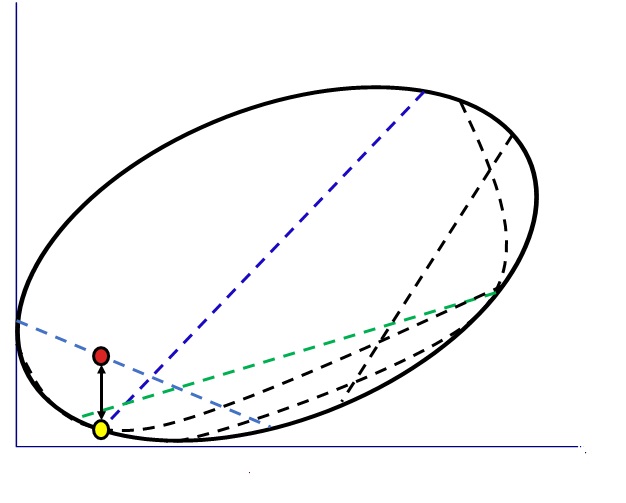}\label{Conv3}}  \ \ \subfloat[]{\includegraphics[width=0.15\textwidth]{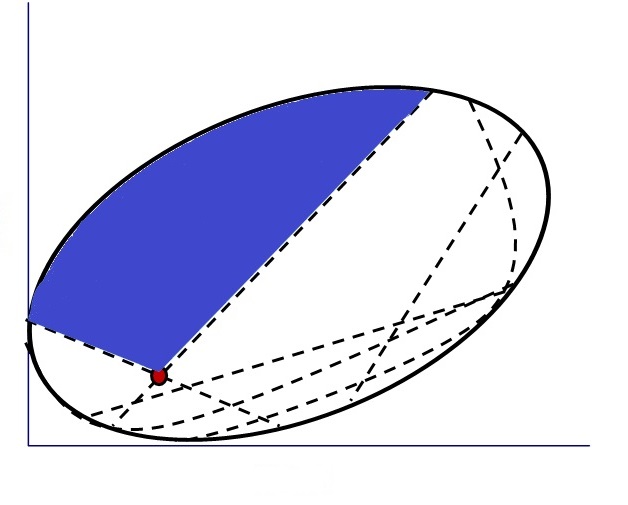}\label{Conv4}}\\
      \captionsetup{justification=raggedright,singlelinecheck=false}
  \caption{\footnotesize Two-dimensional example of algorithm 1: a) normal operation b) adding the first worst-case scenario c) adding the second worst-case scenario d) final feasible region after two iteration.} 
  \label{fig_Conv}
\end{figure}

In this section, we propose the CRSFT method to reduce the size of our optimization problem. CRSFT is an iterative approach with two steps. The first step solves a multi-period OPWF problem of the EWN by considering the regular operation constraints and the crucial constraints. The feasibility check of the contingencies is investigated in the second step. All contingencies are considered as a controllable contingency (CC), and UCs set will be empty in the first iteration. The optimal values for decision variables are obtained by solving the multi-period OPWF problem in a normal situation. These optimal values are used in the second step to check the feasibility of CCs simultaneously to accelerate reaching the SCOEWN. To obtain the CCs for the next iteration, we update constraints (\ref{eq_SCOPF_1}), (\ref{eq_SCOPF_2}), and (\ref{eqSW_1}) with constraints (\ref{eq_iter_1}), (\ref{eq_iter_2}), and (\ref{eq_iter_3}), respectively:
	\begin{align}
&\resizebox{.882\hsize}{!}{$P_{i,t}^\mathrm{g}+{\Delta P}_{i,t}^{\mathrm{g},k}\!-\!P_{i,t}^\mathrm{l}\! + \!{\hat{P}}_{i,t}^{\mathrm{lp},k}\! - \!{\hat{P}}_{i,t}^{\mathrm{ln},k} \!=\!r_{ij}\mathcal{I}_{ij,t}^{k}\!-\!{P_{ji,t}^{k}}\!+\!\sum_{m} \!{P_{ki,t}^{k}}$},\label{eq_iter_1}\\ 
&\resizebox{.882\hsize}{!}{$Q_{i,t}^\mathrm{g}\!\!+\!\!{\Delta Q}_{i,t}^{\mathrm{g},k}\!\!-\!\!Q_{i,t}^\mathrm{l}\! +\! {\hat{Q}}_{i,t}^{\mathrm{lp},k}\! -\! {\hat{Q}}_{i,t}^{\mathrm{ln},k}\! =\!  \!x_{ij}\mathcal{I}_{ij,t}^{k}\!-\!Q_{ji,t}^{k}\!+\!\!\sum_{m}\! Q_{ki,t}^{k}$} \label{eq_iter_2}, \\
&\sum\limits_{l}\!f^{k}_{nl,t} \!=\!   
F^\mathrm{R}_{n,t}\!+\!\Delta F^{\mathrm{R},k}_{n,t}\!- \!d_{n,t}\! -\! {\hat{d}}_{n,t}^{\mathrm{wp},k}\! +\! {\hat{d}}_{n,t}^{\mathrm{wn},k} \!+\!F^{\mathrm{T},k}_{n,t} \label{eq_iter_3},
	\end{align}
where ${\hat{P}}_{i,t}^{\mathrm{lp},k}$, ${\hat{P}}_{i,t}^{\mathrm{ln},k}$, $ {\hat{d}}_{n,t}^{\mathrm{wp},k}$, and ${\hat{d}}_{n,t}^{\mathrm{wn},k}$ are positive variables that show the mismatch loads for power demand and water demand. The following optimization problem is solved for all CCs:
\begin{equation} \label{eq_iter_obj}
\Lambda^k = \mathrm{min} \sum\limits_{i,t}\Big[ {\frac{{\hat{P}}_{i,t}^{\mathrm{lp},k}+{\hat{P}}_{i,t}^{\mathrm{ln},k}}{P^\mathrm{l}}}+\frac{\hat{d}_{n,t}^{\mathrm{wp},k} + {\hat{d}}_{n,t}^{\mathrm{wn},k}}{d}\Big].
\end{equation}
where $P^l$ and $d$ are daily power and water demand, respectively, to provide a comparable situation between the PDN and the WDS. We named the contingencies without any violation ($\Lambda^k=0$) as CCs. The maximum violation between UCs shows the critical contingency that should be added to the master problem for the next iteration. This procedure will be terminated when all of the CCs lead to a feasible solution by taking into account the corrective actions in the second step. Algorithm 1 and Fig. \ref{fig_Conv} show the details and procedure of the CRSFT method. Let $\mathscr{F}$ be the feasible region of normal operations for an EWN, shown with the solid line in Fig \ref{fig_Conv}. Each contingency provides a new feasible region as shown with the dashed line. Let the red dot be the optimal solution of the optimization problem in the first step. Therefore, there are two CCs and four UCs in the set of contingencies. The distance between the red and yellow dots shows the maximum violation. We have used convex relaxation to have the convex feasible region of the regular and secure operation of the EWN. Therefore, adding the worst-case UC to the master problem provides a new convex region, like $\mathscr{F}_c$, with more CCs. The blue dashed line in Fig \ref{Conv2} shows the worst-case UC that should be added to the master problem for the next iteration. This contingency provides a new convex feasible region, $\mathscr{F}_{c1} \subset \mathscr{F}_{c}$. The $\mathscr{F}_{c1}$ consists of four CCs and two UCs. The worst-case UC for the next iteration is shown with the green dashed line in Fig \ref{Conv3}. A new convex set, $\mathscr{F}_{c2} \subset \mathscr{F}_{c1}$,  
is provided by adding this contingency to the master problem. The blue region of Fig \ref{Conv4} shows the feasible region of the operation of the EWN. Since all of the contingencies in the $\mathscr{F}_{c2}$ are CC, the optimal solution of the master problem is SCOEWN.

\begin{algorithm}[!t]
\captionsetup{font={small,sf,bf}, labelsep=newline}
  \caption{\small CRSFT method to obtain the SCOEWN.} 
  \small
\begin{algorithmic}[1] 
\STATE Define the set of contingencies ($C$), initial set of CCs ($C_0^c = C$), initial set of UCs ($C_0^\mathrm{u} = \varnothing$), $j=0$;
\STATE Define $C_{j+1}^c = \varnothing$, $C_{j+1}^\mathrm{u} = \varnothing$; 
\STATE Apply the optimization solver to find the optimal value of decision variables ($P^\mathrm{g}_{i,t}$, $Q^\mathrm{g}_{i,t}$, $F^\mathrm{R}_{n,t}$ and $\beta_{nl,t}$) subject to the network constraints in the normal operation and $C_j^\mathrm{u}$
\STATE Find the $\Lambda^k$ for  $C_j^\mathrm{c}$ based on (\ref{eq_iter_obj})
\STATE Determine the worst-case contingency ($c^\mathrm{w}_j$), which is the contingency with maximum value of $\Lambda^k$ ($\Lambda^\mathrm{w}_j$)
\IF {$\Lambda^\mathrm{w}_j \geq \epsilon$}
\STATE $C_{j+1}^\mathrm{u}\gets C_{j}^\mathrm{u} + c^\mathrm{w}_j$
\STATE $C_{j+1}^\mathrm{c}\gets C_j^\mathrm{c}-c^\mathrm{w}_j$
\STATE $j=j+1$
\STATE go to step \textbf{2} 
\ELSE{}
\STATE The optimal values in step \textbf{3} guarantee SCOEWN.
\ENDIF
\end{algorithmic}
\end{algorithm}
\section{Case Studies}  \label{sec:casestudy}
Based upon the different characterization of the PDN and WDS in different areas \cite{gilvanejad2021introduction}, we present two case studies to demonstrate the robustness and effectiveness of the proposed method. A 24-hour nodal price and the load curve of power demands \cite{narang2011high} are used for both case studies. All simulations are executed in MATLAB
R2019b environment with Intel (R) Core (TM) i7-9700 CPU 3
GHz and 16 GB RAM based personal PC.

\subsection{IEEE 13-Bus PDN with 8-Node EPANET WDS}
\begin{figure}[!htb]
    \vspace{-.2cm}
  \centering
\includegraphics[width=0.375\textwidth]{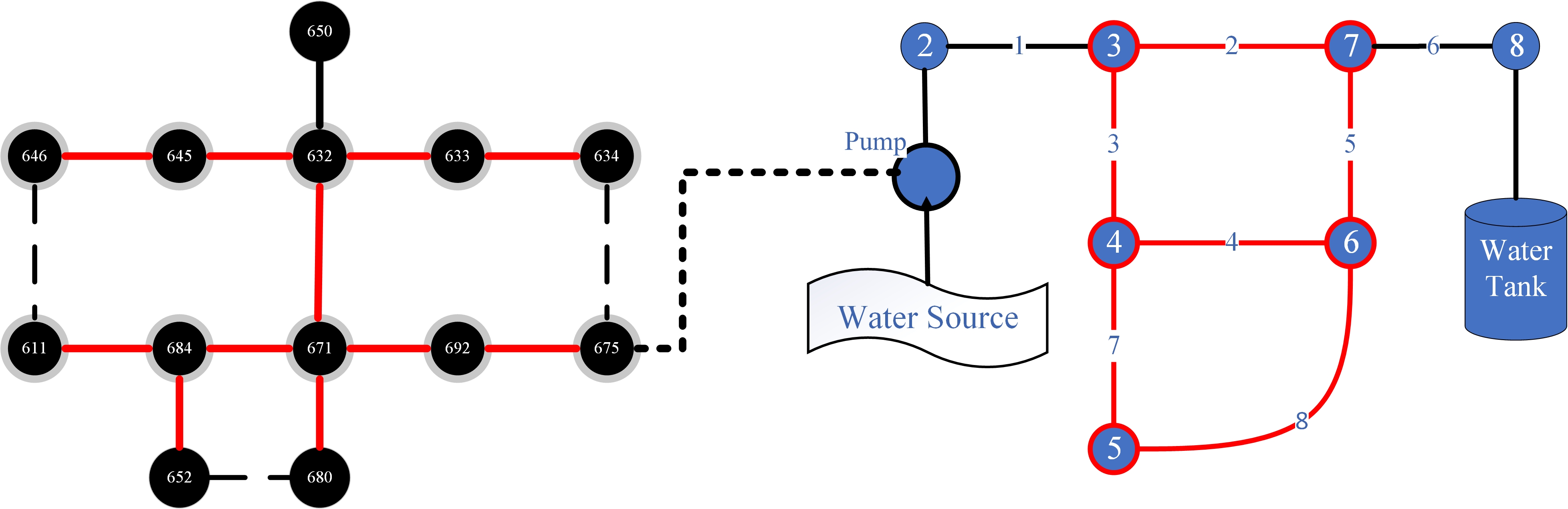}
 \centering
\caption{\footnotesize IEEE 13-bus and EPANET 8-node.}
  \label{EPANET_IEEE}  
        \vspace{-.5cm}
\end{figure}

The first case study is a modified IEEE 13-bus system with the 8 nodes EPANET water system \cite{rossman2000epanet} that is suitable for a micro EWN in a small community. We consider 17 different contingencies for power lines and water pipes that are shown with red color in Fig. \ref{EPANET_IEEE}. Dashed lines in the PDN are reserved lines that are used for DNRC in a contingency situation. We have applied the CRSFT method to the first case study. The optimal values for power generation, water production, and OPS in the base-case operation are obtained in 23.64 seconds. Then we have used these optimal values as parameters in the second step. The numerical results are explained in the following subsections.

\subsubsection{Contingency set in the PDN}
The optimal values of the first step are used to find the UCs in the second step. All of the contingencies are UC in the first iteration, and the critical contingency is contingency 3. We add this contingency to the master problem to find the optimal value of decision variables for the second iteration. The optimization solver reaches the results after 36.87 seconds for the first step in the second iteration. After the second iteration, all contingencies have been converted into CC. Therefore, the EWN operator can handle any contingencies while corrective actions are taken into account. The feasibility of the contingencies can be evaluated independently and in parallel. Table \ref{EPANET_IEEE13water} shows the results. As can be seen, in the case of parallel computing, the first and second iterations take a much smaller time.


\subsubsection{Contingency set in WDS}
In this section, we have studied the set of contingencies in the WDS. Four contingencies are determined as UCs, and the critical contingency is related to contingency 2. Therefore, we add contingency 2 to the master problem for the second iteration with a CPU time of 64.52 seconds. The contingency check for the second iteration shows that all of the contingencies are controllable.  Table \ref{EPANET_IEEE13water} shows the results of different iteration for this case study.  Based on Table \ref{EPANET_IEEE13water}, parallel computing can reduce the time taken during the first and second iterations. 


\begin{table}[!htb]
  \vspace{-.2cm}
 \centering
\footnotesize
\captionsetup{labelsep=space,font={footnotesize,sc}}
\caption{First Case Study: Power and Water Contingencies}
\label{EPANET_IEEE13water}
\begin{tabular}{c?cc?cc}
\hline\hline
 & \multicolumn{2}{c?}{\textbf{PDN}} & \multicolumn{2}{c}{\textbf{WDS}} \\ \hline\hline
Iteration  & \multicolumn{1}{c|}{\tabincell{c}{First}}             &   \tabincell{c}{Second}          & \multicolumn{1}{c|}{\tabincell{c}{First}}           &   \tabincell{c}{Second}        \\ \hline
\tabincell{c}{Parallel\\ Time (s)} & \multicolumn{1}{c|}{1.62}             &   1.58          & \multicolumn{1}{c|}{0.51}           &    0.50       \\ \hline
\tabincell{c}{Total\\ Time (s)} & \multicolumn{1}{c|}{5.44}            &    5.08         & \multicolumn{1}{c|}{2.56 }           &    2.52       \\ \hline
\tabincell{c}{UCs} & \multicolumn{1}{c|}{2 to 12}            &    -         & \multicolumn{1}{c|}{2, 3, 5}           &    -       \\ \hline
\tabincell{c}{Selected\\ Contingency} & \multicolumn{1}{c|}{3}            &    -         & \multicolumn{1}{c|}{2}           &    -       \\ \hline\hline
\end{tabular}
       \vspace{-.1cm}
\end{table}

\subsection{IEEE 33-Bus PDN with 13-Node Otsfeld WDS}
\begin{figure}[!htb] 
  \centering
  \subfloat[]{\includegraphics[width=0.26\textwidth]{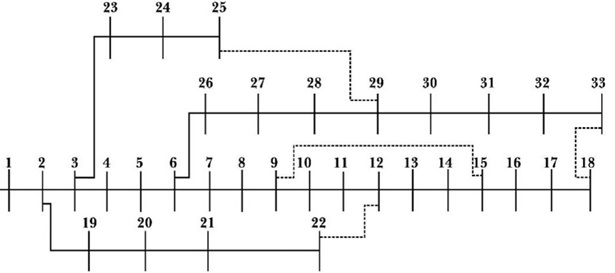}\label{IEEE33bus}}
  \subfloat[]{\includegraphics[width=0.21\textwidth]{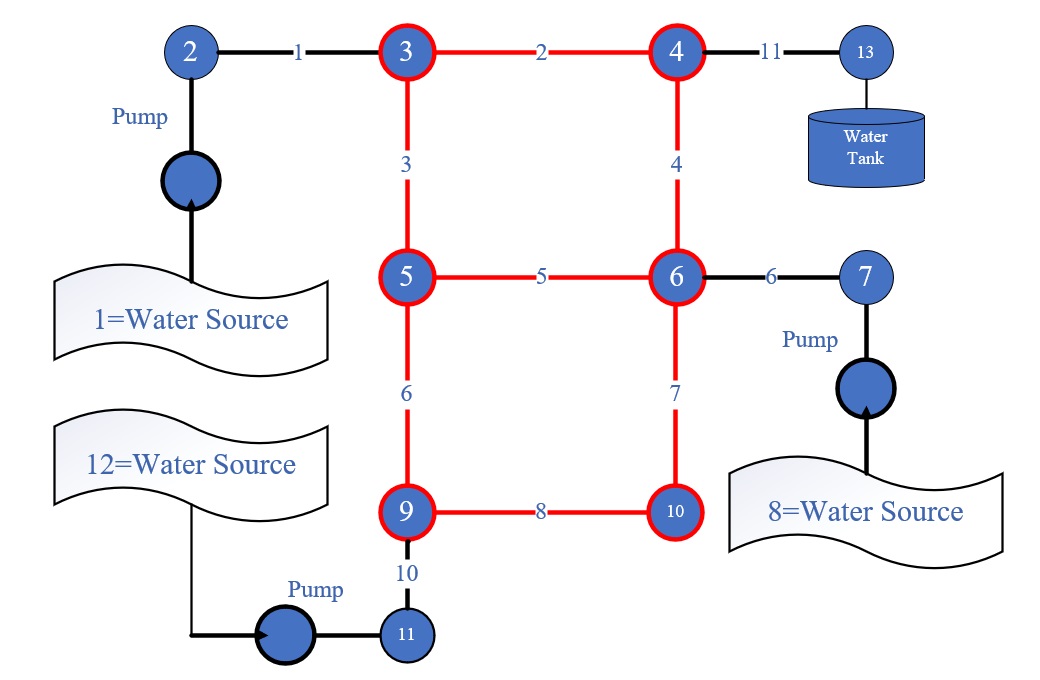}\label{13nodeW}}
\caption{\footnotesize Second test bed: a) IEEE 33-bus system b) Otsfeld water system. }
  \label{VESUr}
         \vspace{-.5cm}
\end{figure}

The IEEE 33-bus system and the Otsfeld regional WDS \cite{ostfeld2011chemical} shown in Fig. \ref{VESUr} are considered for the second case study that can be used for an area of a city. Three pumps that link the PDN to the WDS are connected to buses 14, 24, and 33. The WDS consists of three water resources with three pump stations, 13 nodes, 14 pipes, two PRVs (on pipes 6 and 10), and one water tank. Same as the previous case study, the optimal value for decision variables should be found in normal operation as mentioned in Algorithm 1. The optimization solver reaches the results with a CPU time of 47.66 seconds. These optimal values are used as parameters for the feasibility check.

\subsubsection{Contingency set in PDN}
In the first iteration, all contingencies are UC as a result of solving the optimization problem \ref{eq_iter_obj}. Contingency 31, the worst-case contingency in the first iteration, is added to the master problem for the second iteration. In the second iteration, the master problem is solved in 154.31 seconds, and the updated optimal values of decision variables are used for contingency checks. The results show that the operator can handle any contingency by taking corrective actions. Table \ref{Ostfel_IEEE33} shows the results of different iterations for the second testbed. The total time for contingency checks in the first and second iterations is 8.26 seconds and 7.17 seconds, respectively. By contrast, it takes at least 90.12 seconds and 81.73 seconds without parallel computing.



\subsubsection{Contingency set in WDS}
This section validates the effectiveness of the proposed method for a set of WDS contingencies in the second case study. In the first iteration, contingency 2, contingency 3, contingency 4, and contingency 6 are UCs. All contingencies will be CC in the second iteration by adding contingency 2 to the master problem. The optimization solver finds the optimal values of decision variables for the second iteration in 87.54 seconds. Table \ref{Ostfel_IEEE33} shows the results of different iterations. In parallel computing mode,  the total time for the first and second iterations is 0.24 seconds and 0.20 seconds, respectively. However, without parallel solving, the total time for the first and second iteration will be 1.65 seconds and 1.26 seconds, respectively.


\begin{table}[!htb]
    \vspace{-.2cm}
 \centering
\footnotesize
\renewcommand{\arraystretch}{0.8}
\captionsetup{labelsep=space,font={footnotesize,sc}}
\caption{Second Case Study: Power and Water Contingencies}
\label{Ostfel_IEEE33}
\begin{tabular}{c?cc?cc}
\hline\hline
 & \multicolumn{2}{c?}{\textbf{PDN}} & \multicolumn{2}{c}{\textbf{WDS}} \\ \hline\hline
Iteration  & \multicolumn{1}{c|}{\tabincell{c}{First}}             &   \tabincell{c}{Second}          & \multicolumn{1}{c|}{\tabincell{c}{First}}           &   \tabincell{c}{Second}        \\ \hline
\tabincell{c}{Parallel\\ Time (s)} & \multicolumn{1}{c|}{8.26}             &   7.17          & \multicolumn{1}{c|}{0.2}           &    1.65       \\ \hline
\tabincell{c}{Total\\ Time (s)} & \multicolumn{1}{c|}{90.12}            &    81.73         & \multicolumn{1}{c|}{0.24 }           &    1.26       \\ \hline
\tabincell{c}{UCs} & \multicolumn{1}{c|}{3 to 32}            &    -         & \multicolumn{1}{c|}{2, 3, 4, 6}           &    -       \\ \hline
\tabincell{c}{Selected\\ Contingency} & \multicolumn{1}{c|}{31}            &    -         & \multicolumn{1}{c|}{4}           &    -       \\ \hline\hline
\end{tabular}
\end{table}

To further validate the proposed method, we will look at the effects of selecting the wrong contingency as the critical contingency in the WDS for the second case study. We have added contingency 2 to the master problem which is not the critical contingency. Although this contingency is changed into a CC, all other contingencies will be UC. Hence, adding one of the contingencies as random to the master problem can increase the number of UCs.

\section{Conclusions}\label{sec:conclusions}
The paper presents a security-constrained optimal operation of the EWN (SCOEWN). First, EWN has been modeled in a contingency situation. Then, we have convexified this model to reduce the computational burden. A two-step iterative algorithm is proposed to obtain SCOEWN. The first step solves the optimization problem for the base-case operation and a set of the critical contingencies. The second step consists of considering the violation of the optimal values for all contingencies in order to obtain a new critical contingency. This contingency is added to the master problem for the next iteration. The process continues until there are no uncontrollable contingencies. The method has been tested in two different case studies. The proposed contingency filtering method allows most of the contingencies to be removed from the master problem since they are redundant. With the proposed method, the SCOEWN can be reached very fast.

\bibliographystyle{IEEEtran}	
\bibliography{Main}

\begin{thebibliography}{10}
\providecommand{\url}[1]{#1}
\csname url@samestyle\endcsname
\providecommand{\newblock}{\relax}
\providecommand{\bibinfo}[2]{#2}
\providecommand{\BIBentrySTDinterwordspacing}{\spaceskip=0pt\relax}
\providecommand{\BIBentryALTinterwordstretchfactor}{4}
\providecommand{\BIBentryALTinterwordspacing}{\spaceskip=\fontdimen2\font plus
\BIBentryALTinterwordstretchfactor\fontdimen3\font minus
  \fontdimen4\font\relax}
\providecommand{\BIBforeignlanguage}[2]{{%
\expandafter\ifx\csname l@#1\endcsname\relax
\typeout{** WARNING: IEEEtran.bst: No hyphenation pattern has been}%
\typeout{** loaded for the language `#1'. Using the pattern for}%
\typeout{** the default language instead.}%
\else
\language=\csname l@#1\endcsname
\fi
#2}}
\providecommand{\BIBdecl}{\relax}
\BIBdecl

\bibitem{li2018modeling}
Q.~Li, S.~Yu, A.~Al-Sumaiti, and K.~Turitsyn, ``Modeling and co-optimization of
  a micro water-energy nexus for smart communities,'' in \emph{2018 IEEE PES
  Innovative Smart Grid Technologies Conference Europe (ISGT-Europe)}.\hskip
  1em plus 0.5em minus 0.4em\relax IEEE, 2018, pp. 1--5.

\bibitem{xavier2019transmission}
A.~S. Xavier, F.~Qiu, and F.~Wang, ``Transmission constraint filtering in
  large-scale security-constrained unit commitment,'' \emph{IEEE Trans. Power
  Syst.}, vol.~34, no.~3, pp. 2457--2460, 2019.

\bibitem{park2018sparse}
B.~Park, J.~Holzer, and C.~Marco, ``Sparse tableau formulation for node-breaker
  representations in security-constrained optimal power flow,'' \emph{IEEE
  Trans. Power Syst.}, vol.~34, no.~1, pp. 637--647, 2018.

\bibitem{ardakani2013identification}
A.~J. Ardakani and F.~Bouffard, ``Identification of umbrella constraints in
  dc-based security-constrained optimal power flow,'' \emph{IEEE Trans. Power
  Syst.}, vol.~28, no.~4, pp. 3924--3934, 2013.

\bibitem{weinhold2020fast}
R.~Weinhold and R.~Mieth, ``Fast security-constrained optimal power flow
  through low-impact and redundancy screening,'' \emph{IEEE Trans. Power
  Syst.}, vol.~35, no.~6, pp. 4574--4584, 2020.

\bibitem{capitanescu2008new}
F.~Capitanescu and L.~Wehenkel, ``A new iterative approach to the corrective
  security-constrained optimal power flow problem,'' \emph{IEEE Trans. Power
  Syst.}, vol.~23, no.~4, pp. 1533--1541, 2008.

\bibitem{jiang2013novel}
Q.~Jiang and K.~Xu, ``A novel iterative contingency filtering approach to
  corrective security-constrained optimal power flow,'' \emph{IEEE Trans. Power
  Syst.}, vol.~29, no.~3, pp. 1099--1109, 2013.

\bibitem{fliscounakis2013contingency}
S.~Fliscounakis, P.~Panciatici, F.~Capitanescu, and L.~Wehenkel, ``Contingency
  ranking with respect to overloads in very large power systems taking into
  account uncertainty, preventive, and corrective actions,'' \emph{IEEE Trans.
  Power Syst.}, vol.~28, no.~4, pp. 4909--4917, 2013.

\bibitem{gao2021benders}
H.~Gao and Z.~Li, ``A benders decomposition based algorithm for steady-state
  dispatch problem in an integrated electricity-gas system,'' \emph{IEEE Trans.
  Power Syst.}, 2021.

\bibitem{baran1989optimal}
M.~E. Baran and F.~F. Wu, ``Optimal capacitor placement on radial distribution
  systems,'' \emph{IEEE Trans. Power Deliv}, vol.~4, no.~1, pp. 725--734, 1989.

\bibitem{fooladivanda2017energy}
D.~Fooladivanda and J.~A. Taylor, ``Energy-optimal pump scheduling and water
  flow,'' \emph{IEEE Trans. Control.}, vol.~5, no.~3, pp. 1016--1026, 2017.

\bibitem{goodarzi2021fast}
\BIBentryALTinterwordspacing
M.~Goodarzi, D.~Wu, and Q.~Li, ``Fast security evaluation for operation of
  water distribution systems against extreme conditions,'' 2021. [Online].
  Available: \url{https://arxiv.org/abs/2103.02146}
\BIBentrySTDinterwordspacing

\bibitem{ulanicki2008modeling}
B.~Ulanicki, J.~Kahler, and B.~Coulbeck, ``Modeling the efficiency and power
  characteristics of a pump group,'' \emph{J. Water Resour}, vol. 134, no.~1,
  pp. 88--93, 2008.

\bibitem{li2018micro}
Q.~Li, S.~Yu, A.~S. Al-Sumaiti, and K.~Turitsyn, ``Micro water--energy nexus:
  Optimal demand-side management and quasi-convex hull relaxation,'' \emph{IEEE
  Trans. Control.}, vol.~6, no.~4, pp. 1313--1322, 2018.

\bibitem{gilvanejad2021introduction}
M.~Gilvanejad, M.~Goodarzi, and H.~Ghadiri, ``Introduction of configurational
  indicators for distribution network optimality based on a zoning
  methodology,'' \emph{J. Electr. Eng.}, 2021.

\bibitem{narang2011high}
D.~Narang and C.~Neuman, ``High penetration of photovoltaic generation
  study--flagstaff community power: Results of phase 1,'' DOE Final Technical
  Report, de-ee0002060, Tech. Rep., 2011.

\bibitem{rossman2000epanet}
L.~A. Rossman \emph{et~al.}, ``Epanet 2: users manual,'' \emph{US Environmental
  Protection Agency. Office of Research and Development~…}, 2000.

\bibitem{ostfeld2011chemical}
A.~Ostfeld, E.~Salomons, and O.~Lahav, ``Chemical water stability in optimal
  operation of water distribution systems with blended desalinated water,''
  \emph{J. Water Resour}, vol. 137, no.~6, pp. 531--541, 2011.

\end{thebibliography}

\end{document}